# TRANSMISSION LINE THEORY OF COLLECTIVE PLASMA EXCITATIONS IN PERIODIC TWO-DIMENSIONAL ELECTRON SYSTEMS: FINITE PLASMONIC CRYSTALS AND TAMM STATES

Gregory R. Aizin[a,*] and Gregory C. Dyer[b,†]

[a]*Kingsborough College, The City University of New York, Brooklyn, New York 11235*
[b]*Sandia National Laboratories, P.O. Box 5800, Albuquerque, New Mexico 87185*

We present a comprehensive theory of the one-dimensional plasmonic crystal formed in the grating gated two-dimensional electron gas (2DEG) in semiconductor heterostructures. To describe collective plasma excitations in the 2DEG, we develop a generalized transmission line theoretical formalism consistent with the plasma hydrodynamic model. We then apply this formalism to analyze the plasmonic spectra of 2DEG systems with step-like periodic changes of electron density and/or gate screening. We show that in a periodically modulated 2DEG, a plasmonic crystal is formed and derive closed-form analytical expressions describing its energy band spectrum for both infinite and finite size crystals. Our results demonstrate a non-monotonic dependence of the plasmonic band gap width on the electron density modulation. At so-called transparency points where the plasmon propagates through the periodic 2DEG in a resonant manner, the plasmonic band gaps vanish. In semi-infinite plasmonic crystals, we demonstrate the formation of plasmonic Tamm states and analytically derive their energy dispersion and spatial localization. Finally, we present detailed numerical analysis of the plasmonic band structure of a finite four-period plasmonic crystal terminated either by an Ohmic contact or by an infinite barrier on each side. We trace the evolution of the plasmonic band spectrum, including the Tamm states, with changing electron density modulation and analyze the boundary conditions necessary for formation of the Tamm states. We also analyze interaction between the Tamm states formed at the opposite edges of the short length plasmonic crystal. The validity of our theoretical approach was confirmed in experimental studies of plasmonic crystals in short modulated plasmonic cavities (G. C. Dyer *et. al.*, Phys. Rev. Lett. **109**, 126803 (2012)) which demonstrated excellent quantitative agreement between theory and experiment.

[*] gregory.aizin@kbcc.cuny.edu
[†] gcdyer@sandia.gov

## I. INTRODUCTION

The collective plasma excitations in two-dimensional (2D) electron channels discovered more than 30 years ago [1] have recently returned as a focal point of vigorous research activity. This renewed interest is mostly connected with the problem of the so-called "THz gap" [2] which refers to the historic lack of high power sources and sensitive detectors in the THz band of the electromagnetic (EM) spectrum [2]. The frequencies of 2D plasmons in semiconductor heterostructures conveniently fall in the THz region and can be easily tuned by changing the 2D electron density through an applied field effect. These features make 2D plasmons an excellent candidate to be used for frequency tunable THz detection [3-18]. The predicted instabilities of 2D plasma waves interacting with incident THz EM radiation also provide opportunities for designing THz sources [19-21].

2D plasmons do not interact directly with free space radiation because of the momentum mismatch between THz photons and 2D plasmons. Coupling between free space THz radiation and 2D plasmon is routinely achieved by using a periodic grating gate positioned above the 2D electron channel [1,5-9,17,18]. The grating gate modulates the incident EM wave and defines the plasmon wave vector thus compensating for the momentum mismatch.

The grating gate originally used as a coupler between EM radiation and 2D plasmons in field effect transistors (FETs) also introduces significant changes into the collective electron dynamics in the 2D channel. These changes result from the one-dimensional (1D) periodicity of the various parameters of the 2D electron gas (2DEG) imposed by the grating gate. In particular, the gate screening and equilibrium electron density become periodically modulated with the density modulation being induced by the applied gate bias. As follows from the general Bloch-Floquet theorem in this periodic medium the plasmon spectrum should have an energy band structure, i.e. a 1D plasmonic crystal should be formed.

Plasmonic crystals based upon a 2DEG introduce new possibilities for the design of novel THz devices. Surface plasmonic crystal devices based upon doped semiconductor [22] or metallic [23] periodic structures tend to have weak confinement of the THz field. Also, their plasma frequencies are fixed by the bulk material properties. Planar THz periodic metamaterials with unit cells constructed from bulk materials [24] are similarly constrained by the intrinsic material properties and geometry that define their resonant frequencies and cannot be broadly tuned. Both of these limitations are addressed by 2D plasmonic crystals, which have both tight field confinement around the 2DEG and broad tunability of a resonant electromagnetic medium. It is this field confinement, with scale reduction by several orders of magnitude relative to the free space wavelength, and *in situ* tunability that makes 2D plasmonic devices particularly attractive as tunable passive circuit components [25,26] and direct detectors [12,13] at microwave and THz frequencies.

Plasmonic crystals in the 2DEG were first considered theoretically in the 2D electron system with weak modulation of electron density by Krasheninnikov and Chaplik [27]. They used perturbation theory to calculate minigaps in the 2D plasmon dispersion. The opening of minigaps in the 2D plasmon spectrum has been confirmed experimentally in measurements of the FIR transmission in density modulated Si-MOSFETs [28] and GaAs/AlGaAs heterojunctions [29]. Recently, 2D plasmonic crystal effects have been observed in GaAs/AlGaAs heterostructures with a periodically modulated width of the 2D channel at microwave frequencies. In Ref. [26] a plasmonic crystal band gap has been measured in the energy spectrum of 2D plasmons, and in Ref. [30] the band gaps have been found in the energy spectrum of edge magnetoplasmons. In one of the latest developments, the tunable band spectrum of the finite plasmonic crystal formed in the plasmonic cavity of GaAs/AlGaAs FETs has been measured at sub-THz frequencies [31].

Progress has also been made in theoretical description of plasma excitations in the 2D electron channels with periodically modulated equilibrium electron density. The plasmon energy band spectrum has been found numerically in various approximations [32-36]. Density and field distributions in the split plasmon modes at the edges of the band gap and their interaction with incident EM field have also been analyzed [37-39]. Recently plasmonic band spectrum has been calculated in the gated 2DEG with periodically modulated plasma wave velocity [40]. So far theoretical studies of 2D plasmonic crystals have been mostly concerned with description of the plasma modes at the band gap edges corresponding to the boundaries of the Brillouin zone. These modes are excited when an incident EM wave is coupled to the 2DEG by a grating gate and are probed in transmission and absorption experiments. In the recent experiments an energy spectrum of plasmonic crystal was probed in the plasmonic cavities [31] or waveguides [26] where a variety of plasma modes can be excited. The full energy band structure of the *finite* plasmonic crystal should be used for interpretation of these experiments.

In this paper, we develop a theoretical description of finite plasmonic crystals formed in periodic 2D electron systems. We derive the energy band structure of both infinite and finite 1D plasmonic crystals in the 2DEG with step-like periodic changes of electron density and screening and analyze its evolution with changing density modulation. For the finite plasmonic crystal we also predict formation of the Tamm states at the crystal boundaries and find their energy spectrum and spatial localization. To formulate the plasmonic crystal theory we develop a generalized transmission line (TL) model equivalent to the hydrodynamic approximation of the 2D plasmon theory. This model can be used to describe of 2D plasmons in the long wavelength limit when the plasmon wavelength and phase velocity are larger than the Fermi wavelength and Fermi velocity of 2D electrons respectively. Our theory is also applicable to a broader class of piecewise 2D electron systems in semiconductor nanostructures. Some key elements of the theory have already been used for interpretation of experimental data in our recently published experimental study of a spatially inhomogeneous 2D plasmonic cavity[31].

The rest of this paper is organized as follows. In Section II we develop the TL model and formulate 2D plasmon theory in terms of this model. In Section III the TL model is used to describe plasmons in the general case of a piecewise 2D electron system. These general results are used to derive and analyze the plasmon dispersion equation in the finite as well as infinite plasmonic crystal. The same method is used to describe the Tamm states formed at the boundary of a semi-infinite crystal. In Section IV we present detailed numerical analysis of the finite plasmonic crystal placed into a plasmonic cavity. This includes the energy band structure, the Tamm states and spatial distributions of voltages in various plasma modes. Our conclusions and a brief summary of the results are presented in Section V. The details of some calculations are contained in Appendixes A and B.

## II. TRANSMISSION LINE FORMULATION OF THE 2D PLASMON THEORY

In the quasi-static approximation and long wavelength limit, collective plasma excitations in the 2DEG can be described by the hydrodynamic model which includes the Euler and continuity equations together with the Poisson equation for the self-consistent electric potential $\varphi(\vec{r}, t)$ of the plasma wave [41]. For a plasmon propagating in the $x$-direction in the 2DEG layer positioned in the plane $z = 0$ as shown in Fig. 1 (a) these equations are:

$$\frac{\partial v}{\partial t} + v \frac{\partial v}{\partial x} = \frac{e}{m^*} \frac{\partial \varphi}{\partial x} - \frac{v}{\tau}, \tag{1}$$

$$\frac{\partial n}{\partial t} + \frac{\partial (n_s v)}{\partial x} = 0, \tag{2}$$

$$\nabla^2_{x,z} \varphi = \frac{4\pi e n}{\varepsilon} \delta(z). \tag{3}$$

Here $n_s = n_0 + n$ where $n_0$ is an equilibrium 2D electron density, $n(x,t)$ and $v(x,t)$ are fluctuations of 2D electron density and average velocity, and $-e$ and $m^*$ are the electron charge and effective mass. A phenomenological damping term is included into the Euler equation, Eq. (1), to account for collisional damping of the plasmon with characteristic relaxation time $\tau$. In Eq. (3) it is assumed that the 2DEG is embedded into a medium with dielectric constant $\varepsilon$.

To find the connection between the electric potential $\varphi$ and the 2D electron charge density fluctuation $\rho = -en$ we assume that an ideal metal gate parallel to the 2D plane is positioned at $z = d$ as shown in Fig. 1(a). Then Eq. (3) must be supplemented by the boundary condition $\partial \varphi(x, z = d, t)/\partial x = 0$. In a plasma plane wave of frequency $\omega$ propagating in the positive $x$-direction, the fluctuations $v_{q\omega}(x,t), \rho_{q\omega}(x,t) \propto exp(-iqx + i\omega t)$, where $q = q' - iq''$ ($q', q'' > 0$) is a complex wave vector. In this case, solving Eq. (3) with gate boundary conditions we find

$$\varphi_{q\omega}(x, z, t) = \frac{4\pi \rho_{q\omega}(x,t) e^{-qd}}{\varepsilon q} \begin{cases} e^{qz} \sinh qd, & z < 0 \\ \sinh q(d-z), & 0 < z < d \\ 0, & z > d \end{cases}. \tag{4}$$

For small fluctuations $n_{q\omega}$ and $v_{q\omega}$ linearized Eqs. (1) and (2) together with Eq. (4) yield the following two equations

$$\frac{\partial j_{q\omega}(x,t)}{\partial t} = -\frac{e^2 n_0}{m^*} \frac{\partial \varphi_{q\omega}(x, z=0, t)}{\partial x} - \frac{j_{q\omega}(x,t)}{\tau}, \tag{5}$$

$$\frac{\varepsilon q(1+\coth qd)}{4\pi}\frac{\partial \varphi_{q\omega}(x,z=0,t)}{\partial t} + \frac{\partial j_{q\omega}(x,t)}{\partial x} = 0 , \tag{6}$$

where $j_{q\omega} = -en_0 v_{q\omega}$ is 2D current density fluctuation. For plane wave fluctuations $j_{q\omega}$ and $\varphi_{q\omega}$ it follows from Eqs. (5) and (6) that a non-trivial solution is possible only if

$$\omega\left(\omega - \frac{i}{\tau}\right) = \frac{4\pi e^2 n_0 q}{m^*\varepsilon(1+\coth qd)} . \tag{7}$$

The last equation represents the well-known 2D plasmon dispersion law in the gated 2D electron system [42]. For a 2D plasma wave of frequency $\omega$ it determines the complex wave vector $q$.

To develop a transmission line (TL) description of the 2D plasma wave of frequency $\omega$, we rewrite the hydrodynamic equations (5) and (6) as follows

$$\frac{\partial V_\omega(x,t)}{\partial x} = -L\frac{\partial I_\omega(x,t)}{\partial t} - RI_\omega(x,t), \tag{8}$$

$$\frac{\partial I_\omega(x,t)}{\partial x} = -C(\omega)\frac{\partial V_\omega(x,t)}{\partial t} , \tag{9}$$

$$R = \frac{m^*}{e^2 n_0 \tau W} , \quad L = \frac{m^*}{e^2 n_0 W} , \quad C(\omega) = \frac{W\varepsilon q(1+\coth qd)}{4\pi} . \tag{10}$$

Here $I_\omega(x,t) = I_\omega(x)\exp(i\omega t) \equiv j_{q\omega}(x,t)W$ is the total plasma current in a 2D channel of width $W$ and $V_\omega(x,t) = V_\omega(x)\exp(i\omega t) \equiv \varphi_{q\omega}(x,z=0,t)$ is the voltage along the 2D channel. In these equations and below we omitted $q$-dependence in all formulas because at given $\omega$ wave vector $q$ is uniquely determined by Eq. (7). Equations (8) and (9) are the Telegrapher's Equations of a standard TL theory [43]. The coefficients $R$, $L$ and $C(\omega)$ in Eq. (10) represent distributed resistance, inductance, and capacitance per unit length of the equivalent TL shown in Fig. 1 (b) and allow direct physical interpretation. $R$ and $L$ are resistance and kinetic inductance per unit length of the 2D electron channel in the classical Drude model [25]. The coefficient $C(\omega)$ can be interpreted as an effective electrostatic capacitance per unit length of the gated 2D channel with $\delta\rho_\omega(x,t) = C(\omega)\delta\varphi_\omega(x,z=0,t)$. In the limit $qd \to 0$ (strong gate screening), $C(\omega)$ reduces to the standard local gate capacitance $C = W\varepsilon/4\pi d$. In the opposite limit $qd \to \infty$ (ungated 2D electron channel), we recover the expression $C = W\varepsilon q/2\pi$ first derived in Ref. [44]. In the lossy TL ($q'' \neq 0$), the capacitance $C(\omega)$ has non-zero imaginary part. It can be represented as a shunt conductor in the TL equivalent circuit with conductance $G(\omega) = -\omega\, Im\, C(\omega)$, see Fig. 1 (b). The characteristic impedance of the plasmonic TL is determined as

$$Z_0(\omega) = \frac{V_\omega}{I_\omega} = \sqrt{\frac{R+i\omega L}{i\omega C}} = \frac{4\pi}{\varepsilon\omega W(1+\coth qd)} . \tag{11}$$

Solution of the Telegrapher's Equations (8) and (9) reproduces the plasmon dispersion law in the form

$$q = -i\sqrt{i\omega C(\omega)(R + i\omega L)} \tag{12}$$

equivalent to Eq. (7) as well as spatial distributions of the in-plane electric potential, current and charge density of the plasma wave [31].

Additional complications in the TL formulation of the 2D plasmon theory arise from the non-TEM character of the 2D plasma mode. The electric field of the 2D plasmon has a longitudinal component except for the case of strong gate screening whereas the standard TL model describes the TEM modes. This inconsistency is evident in the calculation of the energy flow by the non-TEM 2D plasma mode [44]. The average complex power carried by the plasmon wave in the $x$-direction can be found as

$$P_\omega(x) = \int_{-\infty}^{\infty} dz \int_{-W/2}^{W/2} dy \langle S_x \rangle , \tag{13}$$

where $\langle S_x \rangle = (c/8\pi)\vec{E} \times \vec{H}^*$ is the Poynting vector averaged over the THz period. The integral in Eq. (13) is evaluated in Appendix A to yield

$$P_\omega(x) = \frac{1}{2} V_\omega(x) I_\omega^*(x) \xi^*(\omega, d) , \tag{14}$$

where

$$\xi(\omega, d) = 1 - \frac{q(1 - e^{-2q'd} \cos 2q''d)}{2q'(1 - e^{-2q^*d})} + \frac{q e^{-2q'd} \sin 2q''d}{2q''(1 - e^{-2q^*d})} . \tag{15}$$

The result in Eq. (14) differs from standard TL expression $(1/2)V_\omega(x)I_\omega^*(x)$ by a form-factor $\xi^*(\omega, d)$. This problem is known in the theory of non-TEM waveguide systems such as microstrip lines and can be resolved in the following way [45]. The expression for power in Eq. (14) can be reduced to the standard TL form above by defining the effective voltage $\tilde{V}_\omega(x)$ and current $\tilde{I}_\omega(x)$ so as to include the form-factor $\xi(\omega, d)$. $\tilde{V}$ and $\tilde{I}$ are not unique as in the case of standard TL definitions of $V$ and $I$. They are in general some abstract quantities connected to $V$ and $I$. However, $\tilde{V}$ and $\tilde{I}$ should be defined to satisfy $P_\omega(x) = (1/2)\tilde{V}_\omega(x)\tilde{I}_\omega^*(x)$ consistent with Eq. (14). Once one of these quantities is defined in terms of $V$ and $I$ the second one is determined by the power relationship formulated above. As shown in the next section for the plasmonic structures considered in this paper, the most convenient and physically sensible choice is: $\tilde{V}_\omega(x) \equiv V_\omega(x)$ and $\tilde{I}_\omega(x) \equiv I_\omega(x)\xi(\omega, d)$. This is equivalent to redefining the characteristic impedance as

$$\tilde{Z}_0 = \frac{Z_0}{\xi(\omega,d)} . \tag{16}$$

It follows from Eq. (15) that in the case of strong gate screening $\xi(\omega, d) = 1$ as expected for the TEM plasmon mode. In the opposite limit of an unscreened 2D electron layer $\xi(\omega, d) = q^*/2q'$ so that $Re\xi = 1/2$. The latter limit was considered in Ref. [44]. Physically, the form-factor $\xi(\omega, d)$ can be understood as accounting for the contribution of the displacement current due to the longitudinal component of the electric field to the power carried by the plasma wave.

## III. TL THEORY OF PLASMONIC CRYSTALS IN PERIODIC 2D ELECTRON SYSTEMS

In this section we apply the plasmonic TL model developed in Section II to the description of collective excitations in periodic semiconductor heterostructures with a 2D electron gas. Examples of these structures include Si-, GaAs-, and GaN-based FETs with periodic multifinger grating gates [1,5,6,14], Si MOSFETs with a periodically modulated oxide thickness [28], and very recently GaAs/AlGaAs heterostructures with a periodically modulated width of the 2D electron channel [26]. All of these systems feature step-like periodic changes of various parameters of the 2D electron channel such as electron density, screening, and geometric size. These periodically changing parameters enter Eqs. (8) and (9) that define collective plasma excitations in the 2D channel.

As an illustration, a basic design of the grating gated FET with 2D electron channel is shown in Fig. 2 (a). The 2D channel consists of alternating ungated and gated regions of lengths $L_1$ and $L_2$, respectively. In this system, one can periodically modulate an equilibrium electron density by applying constant gate bias. Also, the grating gate produces periodically changing screening of electron fluctuations in the channel. We assume that at the boundary between gated and ungated regions an equilibrium electron density and/or gate screening changes in a step-like manner. This approximation is justified if the effective Bohr radius $a_B$ and the distance between the gate and the 2D channel $d$ are small enough so that $a_B, d \ll L_1, L_2$. These conditions can be easily met experimentally. Then the whole 2D structure can be viewed as a stepped transmission line if proper boundary conditions are set at the interfaces between gated and ungated regions.

As boundary conditions, we assume continuity of the power flow $P_\omega(x)$ defined in Eq. (14) and the voltage $\tilde{V}_\omega(x) = V_\omega(x)$ in the plane of the 2DEG. These boundary conditions imply continuity of the effective plasma current $\tilde{I}_\omega(x)$ but discontinuity of the actual plasma current $I_\omega(x)$. The latter condition is justified if we allow an edge charge accumulation at the boundaries. The edge charge accumulation was confirmed by direct numerical solution of the Maxwell equations in the grating gated 2D electron channel [46]. With these boundary conditions each individual segment of the stepped plasmonic transmission line is represented by the equivalent TL circuit elements defined in Eq. (10) but has characteristic impedance $\tilde{Z}_0$ determined by Eq. (16).

In the grating gated structure shown in Fig. 2 (a), there are two different types of segments corresponding to ungated ($j=1$) and gated ($j=2$) regions as shown on the equivalent plasmonic TL diagram in Fig. 2 (b). The equivalent stepped TL is a two-port network consisting of a series of individual segments. Currents and voltages at the opposite sides of each individual segment are connected by the transfer matrix $\hat{t}_j(\omega)$ [43]:

$$\begin{pmatrix} V_{\omega l}^{(j)} \\ \tilde{I}_{\omega l}^{(j)} \end{pmatrix} = \tilde{t}_j(\omega) \begin{pmatrix} V_{\omega r}^{(j)} \\ \tilde{I}_{\omega r}^{(j)} \end{pmatrix} = \begin{pmatrix} \cos q_j L_j & i\tilde{Z}_{0j} \sin q_j L_j \\ \frac{i}{\tilde{Z}_{0j}} \sin q_j L_j & \cos q_j L_j \end{pmatrix} \begin{pmatrix} V_{\omega r}^{(j)} \\ \tilde{I}_{\omega r}^{(j)} \end{pmatrix}, \qquad (17)$$

where $j = 1,2$, $L_j$ and $\tilde{Z}_{0j}$ are the length and characteristic impedance of the $j^{\text{th}}$ segment and $q_j$ is determined by the dispersion equation, Eq. (12), for a given segment. Indices $r$ and $l$ refer to the right- and left hand sides of the segment, respectively, see Fig. 2 (b). The currents $\tilde{I}_{r(l)}$ and voltages $V_{r(l)}$ at the right (left) termination points of the whole structure are connected by equation

$$\begin{pmatrix} V_l \\ \tilde{I}_l \end{pmatrix} = \begin{pmatrix} T_{11} & T_{12} \\ T_{21} & T_{22} \end{pmatrix} \begin{pmatrix} V_r \\ \tilde{I}_r \end{pmatrix}, \tag{18}$$

where the total transfer matrix $\hat{T}(\omega)$ is the product of the segment matrices $\hat{t}_j(\omega)$ ordered in a sequence determined by the geometry of the structure. Plasmon dispersion law in this structure depends on the boundary conditions at termination points. If the external load impedance at the right (left) termination point is equal to $Z_{r(l)}$ we have

$$V_{r(l)} = \tilde{Z}_{r(l)} \tilde{I}_{r(l)}, \tag{19}$$

where $\tilde{Z}_{r(l)} = Z_{r(l)}/\xi_{r(l)}$ and $\xi_{r(l)}$ is the form-factor of the right (left) terminating segment determined by Eq. (15). Combining Eqs. (18) and (19) we obtain

$$\frac{T_{11}(\omega)\tilde{Z}_r + T_{12}(\omega)}{T_{21}(\omega)\tilde{Z}_r + T_{22}(\omega)} = \tilde{Z}_l. \tag{20}$$

The last equation is identical to the expression for the input impedance at the left port of a two-port network with a load impedance of $\tilde{Z}_r$ at its right port [43]. Here, however, the terminal impedances are fixed by the boundary conditions and Eq. (20) thus defines the plasmon dispersion law of the stepped plasmonic TL.

In the subsequent sections, we restrict our consideration of the plasmon dispersion law to the limiting cases of zero and infinite terminal impedances of a stepped plasmonic TL. For example, in a FET structure an Ohmic contact to the 2D channel can be modeled by zero load impedance whereas a terminating region of fully depleted 2DEG can be modeled by infinite load impedance [19]. This latter condition can be achieved by biasing an individual gate finger well beyond pinch-off to induce a barrier in the 2D electron channel. If $Z_l = Z_r = 0$ (two terminating Ohmic contacts) or $Z_l = Z_r = \infty$ (two terminating infinite barriers) Eq. (20) yields the dispersion equation $T_{12}(\omega) = 0$ or $T_{21}(\omega) = 0$, respectively. If $Z_l = 0$, $Z_r = \infty$ (terminating Ohmic contact and infinite barrier) the dispersion equation is $T_{11}(\omega) = 0$. After the plasmon spectrum is found the spatial distributions of the voltage $V_\omega(x)$ and current $I_\omega(x)$ in the plasma wave can be determined in a straightforward way by introducing $x$-dependent $t$-matrices in Eq. (17).

The method presented here is applicable to any segmented 2DEG plasmonic structure with step-like changes of the plasmonic medium properties at the segment boundaries. Below we consider periodic plasmonic structures where a plasmonic crystal is formed provided that the plasmon scattering length is larger than the sample size. Detailed numerical results and analysis of the finite plasmonic crystals are presented in Section IV.

## A. Energy spectrum of an infinite 1D plasmonic crystal

It is instructive to derive the plasmonic dispersion law in a translationally invariant periodic plasmonic medium where an infinite plasmonic crystal is formed. We restrict our consideration to a system like that depicted in Fig. 2 with two different segments in the elementary cell of the plasmonic lattice with lattice constant $L = L_1 + L_2$. We also neglect plasmon damping assuming $\tau \to \infty$ and $R = 0$ in Eq. (12).

Currents and voltages at the opposite sides of elementary cell positioned at $x = 0$ and $x = L$ are connected by the Bloch condition

$$\begin{pmatrix} V_\omega(L) \\ \tilde{I}_\omega(L) \end{pmatrix} = \begin{pmatrix} V_\omega(0) \\ \tilde{I}_\omega(0) \end{pmatrix} e^{ikL}, \tag{21}$$

where $k \in [-\pi/L, \pi/L]$ is plasmon Bloch wave vector. From the TL model we also have

$$\begin{pmatrix} V_\omega(0) \\ \tilde{I}_\omega(0) \end{pmatrix} = \hat{t}_2 \hat{t}_1 \begin{pmatrix} V_\omega(L) \\ \tilde{I}_\omega(L) \end{pmatrix}, \tag{22}$$

where $\hat{t}_2 \hat{t}_1$ is a transfer matrix of the elementary cell with the transfer matrices of individual segments $\hat{t}_j$, $j = 1,2$ defined in Eq. (17). Combining Eqs. (21) and (22) and using the explicit expressions for matrices $\hat{t}_j$ from Eq. (17) we find the dispersion relation for the infinite plasmonic crystal

$$\cos kL = \cos q_1 L_1 \cos q_2 L_2 - \frac{1}{2}\left(\frac{\tilde{Z}_{01}}{\tilde{Z}_{02}} + \frac{\tilde{Z}_{02}}{\tilde{Z}_{01}}\right) \sin q_1 L_1 \sin q_2 L_2, \tag{23}$$

where $q_j$, $j = 1,2$ are defined in Eq. (12). Equation (23) is a particular case of the well-known general equation of the Kronig-Penney model [47] also used to describe the energy spectrum of 1D electron and photonic crystals [48]. Numerical analysis of Eq. (23) is presented in Section IV.

In Ref. [40], the dispersion equation for an infinite 1D plasmonic crystal similar to Eq. (23) was derived in the hydrodynamic approximation in the special case that both segments in a crystal elementary cell are strongly screened by the gate but have either different equilibrium electron densities or different gate-to-channel distances $d$. In the limit of strong gate screening ($qd \ll 1$) and $\tau \to \infty$ Eq. (7) yields the linear plasmon dispersion law [42]

$$\omega = v_p q, \quad v_p = \sqrt{\frac{4\pi e^2 n_0 d}{m^* \varepsilon}}, \tag{24}$$

where $v_p$ is the plasma wave velocity. In the same limit Eqs. (11), (15), and (16) yield

$$\tilde{Z}_0 = \frac{4\pi d}{\varepsilon W v_p}. \tag{25}$$

Combining Eqs. (24) and (25) with Eq. (23) we obtain

$$\cos kL = \cos\frac{\omega L_1}{v_{p1}}\cos\frac{\omega L_2}{v_{p2}} - \frac{1}{2}\left(\frac{d_1 v_{p2}}{d_2 v_{p1}} + \frac{d_2 v_{p1}}{d_1 v_{p2}}\right)\sin\frac{\omega L_1}{v_{p1}}\sin\frac{\omega L_2}{v_{p2}}. \tag{26}$$

The last equation coincides with the dispersion relation found in Ref. [40] only when $d_1 = d_2$.

## B. Tamm states in a semi-infinite 1D plasmonic crystal

If the crystal lattice is terminated, the broken translational symmetry results in the localization of the Bloch waves near the termination point. These states known as the Tamm states [49] have been thoroughly studied theoretically [50] and observed experimentally in electron superlattices [51] and photonic crystals [52]. Below, we analyze the Tamm states in plasmonic crystal and find their spectrum and localization length.

The geometry of the semi-infinite plasmonic lattice is shown in Fig. 2 (c) where the lattice is terminated by the impedance $Z_T$ at $x = -L_1$. In this geometry, the general solution of the Telegrapher's Equations, Eqs. (8) and (9), is [43]:

$$\begin{cases} V(x) = Ae^{-iq_1 x} + Be^{iq_1 x} \\ \tilde{I}(x) = \frac{A}{\tilde{Z}_{01}}e^{-iq_1 x} - \frac{B}{\tilde{Z}_{01}}e^{iq_1 x} \end{cases} \quad -L_1 \leq x \leq 0,$$

$$\begin{cases} V(x) = Ce^{-iq_2 x} + De^{iq_2 x} \\ \tilde{I}(x) = \frac{C}{\tilde{Z}_{02}}e^{-iq_2 x} - \frac{D}{\tilde{Z}_{02}}e^{iq_2 x} \end{cases} \quad 0 \leq x \leq L_2. \tag{27}$$

It follows from the Bloch condition that

$$\begin{cases} V(x) = e^{ikL}\left(Ae^{-iq_1(x-L)} + Be^{iq_1(x-L)}\right) \\ \tilde{I}(x) = e^{ikL}\left(\frac{A}{\tilde{Z}_{01}}e^{-iq_1(x-L)} - \frac{B}{\tilde{Z}_{01}}e^{iq_1(x-L)}\right) \end{cases} \quad L_2 \leq x \leq L. \tag{28}$$

Continuity of $V(x)$ and $\tilde{I}(x)$ at $x = 0$ and $x = L_2$ yields the system of four linear homogeneous equations for the constant coefficients $A$, $B$, $C$, and $D$. This system has non-trivial solution if the dispersion equation, Eq. (23), is satisfied. In this case, the coefficients $B$, $C$, and $D$ can be written in terms of a single coefficient $A$. In particular,

$$B = \frac{\cos q_2 L_2 - \frac{i\tilde{Z}_{02}}{\tilde{Z}_{01}}\sin q_2 L_2 - e^{ikL}e^{iq_1 L_1}}{e^{ikL}e^{-iq_1 L_1} - \cos q_2 L_2 - \frac{i\tilde{Z}_{02}}{\tilde{Z}_{01}}\sin q_2 L_2} A. \tag{29}$$

Voltage and current at the termination point $x = -L_1$ can be found from Eqs. (27) and (29):

$$V(-L_1) = 2iA \frac{\sin q_1 L_1 \cos q_2 L_2 + \frac{\tilde{Z}_{02}}{\tilde{Z}_{01}} \cos q_1 L_1 \sin q_2 L_2}{\cos q_2 L_2 + \frac{i\tilde{Z}_{02}}{\tilde{Z}_{01}} \sin q_2 L_2 - e^{ikL} e^{-iq_1 L_1}}, \tag{30a}$$

$$\tilde{I}(-L_1) = \frac{2A}{\tilde{Z}_{01}} \frac{\cos q_1 L_1 \cos q_2 L_2 - \frac{\tilde{Z}_{02}}{\tilde{Z}_{01}} \sin q_1 L_1 \sin q_2 L_2 - e^{ikL}}{\cos q_2 L_2 + \frac{i\tilde{Z}_{02}}{\tilde{Z}_{01}} \sin q_2 L_2 - e^{ikL} e^{-iq_1 L_1}}. \tag{30b}$$

At the termination point the current and voltage are connected by Eq. (19)

$$V(-L_1) = \tilde{Z}_T \tilde{I}(-L_1). \tag{19a}$$

Equations (23), (30), and (19a) determine the energy spectrum and spatial localization of the plasmonic Tamm states formed near the termination point. Solutions of these equations for a plasmonic crystal lattice terminated by the Ohmic contact ($\tilde{Z}_T = 0$) or the infinite barrier ($\tilde{Z}_T = \infty$) are presented in Appendix B with the following results.

The Bloch wave vector $k$ of the Tamm states is complex

$$k = \frac{\pi n}{L} + i\mu, \quad n = 0,1,2,\ldots \tag{31}$$

and describes the plasmonic energy states positioned in the $n^{\text{th}}$ energy gap of the band spectrum of the infinite plasmonic crystal [50]. The imaginary part of the wave vector represents an inverse localization length of the plasmonic Tamm state. For geometry shown in Fig. 2 (c) only solutions with $\mu > 0$ should be retained.

If the plasmonic crystal is terminated by an Ohmic contact, then energies of the Tamm states are determined by the equation

$$\sin q_1 L_1 \cos q_2 L_2 + \frac{\tilde{Z}_{02}}{\tilde{Z}_{01}} \cos q_1 L_1 \sin q_2 L_2 = 0, \tag{32a}$$

and its localization length can be found from the equation

$$(-1)^n e^{-\mu L} = \cos q_1 L_1 \cos q_2 L_2 - \frac{\tilde{Z}_{01}}{\tilde{Z}_{02}} \sin q_1 L_1 \sin q_2 L_2. \tag{33a}$$

If the plasmonic crystal is terminated by an infinite barrier, the respective equations are

$$\sin q_1 L_1 \cos q_2 L_2 + \frac{\tilde{Z}_{01}}{\tilde{Z}_{02}} \cos q_1 L_1 \sin q_2 L_2 = 0, \tag{32b}$$

$$(-1)^n e^{-\mu L} = \cos q_1 L_1 \cos q_2 L_2 - \frac{\tilde{Z}_{02}}{\tilde{Z}_{01}} \sin q_1 L_1 \sin q_2 L_2. \tag{33b}$$

It follows from Eqs. (32) and (33) that in the semi-infinite crystal terminated by the infinite barrier (Ohmic contact) condition $\mu > 0$ can be satisfied only if $\tilde{Z}_{01} > \tilde{Z}_{02}$ ($\tilde{Z}_{01} < \tilde{Z}_{02}$). We present numerical analysis of the Tamm states in the next section.

## IV. RESULTS AND DISCUSSION

In this section we apply the theoretical formalism developed in Section III to analyze numerically the energy band structure and the Tamm states in a plasmonic crystal formed in the grating gated 2D electron system shown in Fig. 2 (a). In all of the calculations presented below we neglect plasmon damping. We also assume that the lengths of gated and ungated regions are the same, $L_1 = L_2$, and that the gate is positioned at a distance $d = 0.1L$ from the 2D channel (see the system's geometry on Fig. 2 (a)).

First, we analyze the plasmonic band spectrum of the infinite crystal by solving numerically Eq. (23). In this calculation we assume that characteristic impedances of the ungated regions $\tilde{Z}_{01}$ in Fig. 2 (b) are given by Eqs. (11), (15), and (16) taken in the limit $d \to \infty$ and at zero damping, $q'' = Im\, q = 0$, so that $\xi = 1/2$ in Eq. (15) and

$$\tilde{Z}_{01} = \frac{4\pi}{\varepsilon \omega W}. \tag{34}$$

Characteristic impedances of the gated regions are given by the same equations with $q'' = 0$ but $d = 0.1L$. The results are presented in Fig. 3 (a) where the crystal energy band structure is shown as a function of $\gamma = \tilde{n}_0/n_0$ with $\tilde{n}_0$ and $n_0$ being equilibrium electron densities in the gated and ungated regions, respectively. The value of $\gamma$ can be changed experimentally by varying the applied gate bias. All plasma frequencies $\omega$ are normalized to the plasma frequency in the ungated region at wave vector $q = 2\pi/L$

$$\omega_1 = \sqrt{\frac{4\pi^2 e^2 n_0}{\varepsilon m^* L}} \tag{35}$$

which is the fundamental frequency of the size-quantized transverse plasma oscillations in an ungated 2D electron strip of width $L/2$. The grey shaded areas in Fig. 3 (a) represent the energy gaps in the infinite plasmonic crystal spectrum while the non-shaded white areas represent the energy bands. It should be pointed out that the energy gaps in the plasmonic spectrum do not disappear at $\gamma = 1$ when the electron density in the 2D channel is uniform. This happens because the 2DEG is still periodically screened, and $\tilde{Z}_{01} \neq \tilde{Z}_{02}$. In general, the band gap width is a non-monotonic function of $\gamma$. For higher order band gaps there are some values of $\gamma$ where the band gap width is equal to zero. These points can be called the transparency points and have a rather straightforward explanation. The grating gate induced plasmonic lattice in Fig. 2 (a) consists of two sublattices: one comprised of the gated segments and another one comprised of the ungated segments. The energy band spectrum is formed due to coherent coupling of the plasma oscillations in individual segments. The plasmon wave vectors in segments of length $L/2$ are quantized: $q_p = 2\pi p/L$, $p = 1,2,...$. The plasmon eigenfrequencies in individual ungated ($\omega_p$) and gated ($\bar{\omega}_p$) segments are determined by Eq. (7) at $\tau^{-1} = 0$:

$$\omega_p = \sqrt{p}\,\omega_1$$

$$\bar{\omega}_p = \frac{\sqrt{2p\gamma}\omega_1}{\sqrt{1+\coth q_p d}} \qquad p = 1,2,\dots\ , \qquad (36)$$

where $\omega_1$ is defined in Eq. (35). The plasma eigenfrequencies in the gated segments are lower than those in the ungated ones because of the gate screening and lower 2D electron densities. Therefore, at low frequencies only gated sublattice plasma modes are excited as was confirmed experimentally in Ref. [31]. At higher frequencies both gated and ungated sublattice modes are excited producing an entangled and more condensed band structure as shown in Fig. 3 (a). The band gaps disappear when plasma eigenmodes in both gated and ungated segments resonate with each other; that is at $\omega_p = \bar{\omega}_{p'}$ for any integer $p$ and $p'$. In Fig 3 (a) we illustrate these conclusions by plotting the frequencies from Eq. (36) as a function of $\gamma$ with dashed lines. Transparency points correspond to the resonant propagation of the plasma wave through the entire crystal. In Fig. 3 (b) we plot the spatial distribution of the voltage in the plasma wave calculated for the transparency point indicated by an arrow in Fig. 3 (a) to demonstrate this resonant behavior.

We also analyzed the Tamm states for the semi-infinite crystal in this model assuming that the last segment before termination point is an ungated one, as shown in Fig. 2 (c). The Tamm states are described by Eqs. (32) and (33). Since in this model of the plasmonic crystal $\tilde{Z}_{01} > \tilde{Z}_{02}$, the Tamm states are formed only if the semi-infinite crystal is terminated by a barrier, i.e. $\tilde{Z}_T = \infty$ in Fig. 2 (c). Energies of the Tamm states found numerically from Eq. (32b) are shown in Fig. 3 (a) by dotted lines. When the value of $\gamma$ changes these states cross the band gaps as expected [50]. Localization length of the Tamm state $l_{loc} = \mu^{-1}$ found from Eq. (33b) takes the minimum value at the center of the band gap and diverges as the Tamm state approaches the band boundary. This behavior is demonstrated in Fig. 3 (c) where we plot $l_{loc}$ as a function of $\gamma$ for the Tamm state positioned in the second band gap in Fig 3 (a). The voltage distribution in the typical Tamm state near the center of the band gap is shown in the inset in Fig. 3 (c).

As the next step, we considered a finite plasmonic crystal formed in the grating gated 2D electron channel with four grating fingers. The crystal can be terminated either by the Ohmic contact or by the infinite barrier on each side. To make our calculations more relevant to the recent experimental work [31] we assume that 2DEG in the ungated regions is still screened by the neighboring gates and surrounding metallization but 2D electron density can be tuned in the gated regions only. In this model, both $\tilde{Z}_{01}$ and $\tilde{Z}_{02}$ are determined by Eqs. (11), (15), and (16) with $Im\ q = 0$ but only $\tilde{Z}_{02}$ changes when the value of $\gamma$ is varied.

We calculated the transfer matrix $\hat{T}$ of the four-period plasmonic crystal from Eqs. (17) and (18). Because the dispersion relation defined in Eq. (12) is in general a transcendental equation, the plasmon wavevector $q_j$ was found numerically for the segments of 2DEG ($j = 1,2$) forming the crystal. We then numerically solved Eq. (20) to find the plasmon energy spectrum. We present our results for the plasmonic crystal terminated by the Ohmic contact and the infinite barrier on the opposite sides, by the two Ohmic contacts, and by the two infinite barriers in Figs. 4 (a), 5 (a), and 6 (a) respectively. The energy band spectrum of the relevant infinite plasmonic crystal was also found from Eq. (23), and the infinite crystal band gaps are indicated in grey in the Figures. We have also extended the range of $\gamma$ to include the values of $\gamma > 1$ achievable at positive gate biases. There are some noticeable changes in the infinite crystal band structure in this case in comparison to the one shown in Fig. 3 (a). These changes result from the different behavior of the characteristic impedances $\tilde{Z}_{01}$ and $\tilde{Z}_{02}$. In particular, all band gaps vanish at $\gamma = 1$ because at this degeneracy point where $\tilde{Z}_{01} = \tilde{Z}_{02}$, any periodicity in the 2D electron channel disappears and there are no Bragg reflections at the boundary between gated and ungated regions.

The plasmon energies shown in Figs. 4 (a), 5 (a), and 6 (a) by dotted lines tend to bundle in groups of four as expected for the four-period finite crystal. These groups are primarily localized within the bounds of the proper energy bands of the infinite crystal. However, there are also some states positioned in the band gaps of the infinite crystal. These states represent the Tamm states in the finite crystal. Spatial localization of the Tamm states depends on the boundary conditions at the termination points and the value of $\gamma$. At $\gamma < 1$ ($\gamma > 1$) we have $\tilde{Z}_{01} < \tilde{Z}_{02}$ ($\tilde{Z}_{01} > \tilde{Z}_{02}$) so the Tamm states, if any, should be localized near the terminating Ohmic contact (infinite barrier). These qualitative conclusions have been confirmed by direct numerical simulations.

In Fig. 4 (a) the plasmonic energy spectrum is presented for the four-period crystal terminated by an Ohmic contact on one side and a barrier on the opposite side. The spatial distributions of voltage in the plasmonic Tamm state positioned in the first band gap at two different values of $\gamma$ indicated by arrows on Fig. 4 (a) are plotted in Fig. 4 (b). These plots demonstrate that at $\gamma < 1$ the Tamm state is localized near the Ohmic contact whereas at $\gamma > 1$ it is localized near the infinite barrier on the opposite side of the plasmonic crystal.

In Fig. 5 (a) the plasmonic energy spectrum is shown for the four-period crystal terminated by Ohmic contacts on both sides. With these boundary conditions, the Tamm states are not formed if $\gamma > 1$. With $\gamma < 1$, identical Tamm states are formed near both Ohmic contacts. In the finite crystal they interact with each other and split into symmetric and anti-symmetric combinations producing two hybridized Tamm states in the band gap as shown in Fig. 5 (a). This conclusion is further confirmed in Fig. 5 (b) where we plotted the spatial distribution of voltage calculated for these two states at the point indicated by arrows in Fig. 5 (a).

Similar behavior is found for the four-period crystal terminated by infinite barriers on both sides in Fig. 6. In this case, the Tamm states are not formed if $\gamma < 1$. With $\gamma > 1$, two hybridized symmetric and anti-symmetric Tamm states are formed in every band gap as shown in Fig. 6 (a). The spatial distributions of voltage for these hybridized Tamm states at the value of $\gamma$ indicated by arrows in Fig. 6 (a) are plotted in Fig. 6 (b) and confirm our conclusions.

The interaction of the Tamm states formed at the opposing edges of the planar plasmonic crystal exposes an interesting byproduct of the finite crystal. In the semi-infinite crystal, the plasmonic Tamm state is an edge state localized over a few periods of the crystal. However, when the crystal itself is only several periods long, both edges contribute to the spatial distribution of the Tamm states and effectively interact with each other via the hybridized Tamm state. The strength of the interaction depends on the spatial overlap of the Tamm states at opposing edges of the crystal and is dependent upon the ratio of the localization length $l_{loc}$ to the crystal size. The value of $l_{loc}$ can be tuned by an applied gate bias as shown in Fig. 3 (c) providing additional control of the electromagnetic link between the crystal edges. In this paper, the Tamm states were considered in two ideal limits of the boundary conditions at the crystal edge: Ohmic contact with zero impedance and the barrier with infinite impedance. Termination with a finite impedance $\tilde{Z}_T$ changes both the Tamm state energy and its spatial distribution, providing significant versatility in the design of the electromagnetic coupling between the plasmonic crystal edges. This work is in progress.

## V. CONCLUSIONS AND SUMMARY

The results presented in this study have a variety of implications concerning the development of 2DEG based plasmonic crystal devices. Here we have shown the versatility of the plasmonic crystal energy spectrum and the possibility to devise a crystal with required properties by choosing the proper geometry of a structure. The plasmonic spectrum can be tuned *in situ* by varying the gate bias. For typical semiconductor 2DEG densities and readily accessible crystal geometries, the tunable plasmonic spectra naturally span microwave [26] and THz [6] frequencies. Further range in the frequencies accessible in the mid infrared may be achieved as the graphene material system evolves into a viable plasmonic device platform [53,54]. Plasmonic Tamm states in these various material systems have special potential when applied for the field enhancement they can produce at the crystal boundary. Similar to the effect of a plasmonic crystal defect [55], the Tamm state can be harnessed for its strong near-field enhancement in combination with the sub-wavelength nature of the 2D plasmon. For example, the Tamm state could be coupled to an adjacent detection element [56].

In summary, we have developed a description of the collective plasma oscillations in 2D electron systems in terms of the transmission line theory equivalent to their description in the hydrodynamic approximation. Based on this theoretical formalism we have developed a general theory of collective plasma excitations in periodic 2D electron systems such as grating gated FETs. We have shown that collective plasma excitations in this system form a plasmonic crystal and have derived closed-form analytical expressions describing its energy band spectrum. Our results show that the widths of the plasmonic band gaps depend on the 2D electron density modulation and vanish in the so called transparency points where the plasmon propagates through the entire periodic 2D electron system in a resonant manner. In semi-infinite plasmonic crystals we have demonstrated the formation of plasmonic Tamm states and have found analytically their energy dispersion and spatial localization. This general theoretical formalism has been used to analyze the energy band structure of the finite plasmonic crystal most relevant to the current experimental studies of plasmonic effects in grating gated FETs [31]. We have performed numerical simulations of the plasmonic band structure including the Tamm states for the four-period plasmonic crystal terminated either by the Ohmic contact or by the infinite barrier on each side. Such a barrier can be produced by a single gate finger biased beyond its pinch-off voltage with complete depletion of the 2D electron channel underneath the finger. We have traced evolution of the plasmonic band spectrum as a function of the electron density modulation induced by the grating gate voltage and found conditions necessary for formation of Tamm states in the finite plasmonic crystal. Recent experimental studies of the plasmonic crystal energy structure in short modulated plasmonic cavities [31] confirm validity of our theoretical approach and demonstrate excellent quantitative agreement between theory and experiment.

**ACKNOWLEDGMENTS**

The work at Sandia National Laboratories was supported by the DOE Office of Basic Energy Sciences. This work was performed, in part, at the Center for Integrated Nanotechnologies, a U.S. Department of Energy, Office of Basic Energy Sciences user facility. Sandia National Laboratories is a multi-program laboratory managed and operated by Sandia Corporation, a wholly owned subsidiary of Lockheed Martin Corporation, for the U.S. Department of Energy's National Nuclear Security Administration under contract DE-AC04-94AL85000. G.R.A. is thankful for financial support from Sandia National Laboratories.

**APPENDIX A**

To evaluate an integral in Eq. (13) we notice that electric field of the 2D plasma wave of frequency $\omega$ has both transverse, $E_{z,\omega}(x,z,t)$, and longitudinal, $E_{x,\omega}(x,z,t)$, components, whereas magnetic field has only one non-zero component, $H_{y,\omega}(x,z,t)$. These components are determined by the following equations

$$E_{z,\omega} = -\frac{\partial \varphi_\omega}{\partial z}, \quad E_{x,\omega} = -\frac{\partial \varphi_\omega}{\partial x}, \tag{A1}$$

$$-\frac{\partial H_{y,\omega}}{\partial z} = \frac{4\pi}{c} j_\omega \delta(z) + \frac{\varepsilon}{c}\frac{\partial E_{x,\omega}}{\partial t}, \tag{A2}$$

where $\varphi_\omega(x,z,t)$ is defined in Eq. (4). Therefore, the *x*-component of the time-averaged Poynting vector in Eq. (13) is given by

$$\langle S_x \rangle = -\frac{c}{8\pi} E_{z,\omega} H_{y,\omega}^* . \tag{A3}$$

Substituting the expression for $\langle S_x \rangle$ from Eq. (A3) into Eq. (13) and making use of Eq. (1) we obtain after integrating by parts

$$P_\omega(x) = -\frac{cW}{8\pi}\int_{-\infty}^{\infty} \varphi_\omega \frac{\partial H_{y,\omega}}{\partial z} dz . \tag{A4}$$

Then, from Eqs. (A2) and (A4) we find

$$P_\omega(x) = \frac{W}{2}\varphi_\omega(x, z=0) j_\omega^* - \frac{i\varepsilon\omega W}{8\pi}\int_{-\infty}^{\infty} \varphi_\omega(x,z)\frac{\partial \varphi_\omega^*(x,z)}{\partial x} dz . \tag{A5}$$

Integration in the last equation can be performed with $\varphi_\omega$ given by Eq. (4) to yield Eq. (14).

**APPENDIX B**

We are looking for the Bloch-type solutions for the plasma waves localized near the termination point of the plasmonic crystal lattice in Fig. 2 (c), that is solutions with the Bloch vector

$$k = \varsigma + i\mu, \qquad \varsigma, \mu > 0 \quad . \tag{B1}$$

We will neglect the damping effects and assume that the plasmonic lattice is terminated by a purely reactive load $\tilde{Z}_T = i\tilde{Z}_T''$. Substituting Eqs. (30a,b) into Eq. (19) we obtain

$$\cos q_1 L_1 \cos q_2 L_2 - \frac{\tilde{Z}_{02}}{\tilde{Z}_{01}} \sin q_1 L_1 \sin q_2 L_2$$
$$- \frac{\tilde{Z}_{01}}{\tilde{Z}_T''} \left( \sin q_1 L_1 \cos q_2 L_2 + \frac{\tilde{Z}_{02}}{\tilde{Z}_{01}} \cos q_1 L_1 \sin q_2 L_2 \right) - e^{i(\varsigma + i\mu)L} = 0 \quad . \tag{B2}$$

After separation of the real and imaginary parts Eq. (B2) yields

$$\varsigma = \frac{\pi n}{L}, \qquad n = 0,1,2,\dots \quad , \tag{B3}$$

$$(-1)^n e^{-\mu L} = \cos q_1 L_1 \cos q_2 L_2 - \frac{\tilde{Z}_{02}}{\tilde{Z}_{01}} \sin q_1 L_1 \sin q_2 L_2$$
$$- \frac{\tilde{Z}_{01}}{\tilde{Z}_T''} \left( \sin q_1 L_1 \cos q_2 L_2 + \frac{\tilde{Z}_{02}}{\tilde{Z}_{01}} \cos q_1 L_1 \sin q_2 L_2 \right) . \tag{B4}$$

Now, with $k$ given by Eq. (31) dispersion relation in Eq. (23) takes the form

$$(-1)^n \cosh \mu L = \cos q_1 L_1 \cos q_2 L_2 - \frac{1}{2}\left( \frac{\tilde{Z}_{01}}{\tilde{Z}_{02}} + \frac{\tilde{Z}_{02}}{\tilde{Z}_{01}} \right) \sin q_1 L_1 \sin q_2 L_2 \quad . \tag{B5}$$

Excluding $\mu$ from Eqs. (B4) and (B5) we obtain dispersion relation for the Tamm states

$$\sin q_1 L_1 \cos q_2 L_2 + \frac{\tilde{Z}_{01}}{\tilde{Z}_{02}} \cos q_1 L_1 \sin q_2 L_2 + \frac{\tilde{Z}_{01}}{\tilde{Z}_T''} \left( \frac{\tilde{Z}_{02}}{\tilde{Z}_{01}} - \frac{\tilde{Z}_{01}}{\tilde{Z}_{02}} \right) \sin q_1 L_1 \sin q_2 L_2$$
$$+ \frac{\tilde{Z}_{01}^2}{\tilde{Z}_{02}^2} \left( \sin q_1 L_1 \cos q_2 L_2 + \frac{\tilde{Z}_{02}}{\tilde{Z}_{01}} \cos q_1 L_1 \sin q_2 L_2 \right) = 0 \quad . \tag{B6}$$

Energies of the Tamm states can be found from Eq. (B6). Then the localization length $\mu^{-1}$ can be determined from Eq. (B4). Using Eq. (B6) one can rewrite Eq. (B4) in another equivalent form:

$$(-1)^n e^{-\mu L} = \cos q_1 L_1 \cos q_2 L_2 - \frac{\tilde{Z}_{01}}{\tilde{Z}_{02}} \sin q_1 L_1 \sin q_2 L_2$$

$$+ \frac{\tilde{Z}_T''}{\tilde{Z}_{01}} \left( \sin q_1 L_1 \cos q_2 L_2 + \frac{\tilde{Z}_{01}}{\tilde{Z}_{02}} \cos q_1 L_1 \sin q_2 L_2 \right) \ .$$

(B4a)

Equations (32a) and (32b) follow from Eq. (B6) in the limits $\tilde{Z}_T = 0$ and $\tilde{Z}_T = \infty$, respectively, whereas Eqs. (33a) and (33b) follow from Eqs. (B4) and (B4a) taken in the same respective limits.

**FIGURES**

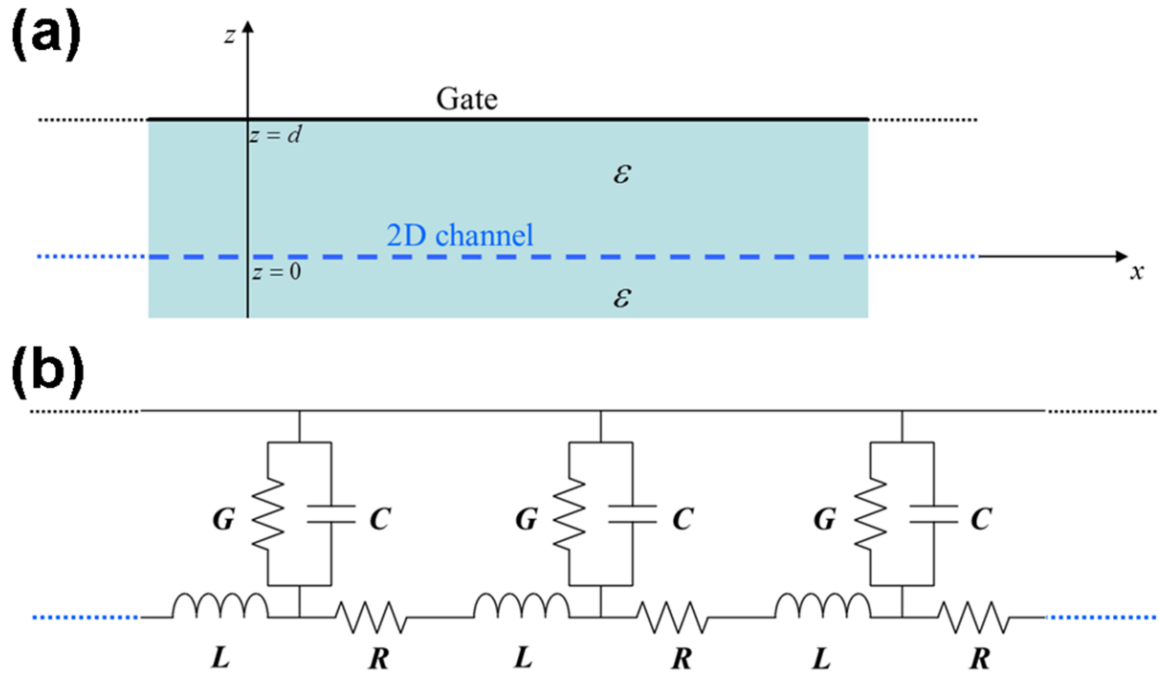

**Figure 1.** (a) Gated 2D electron channel and (b) its equivalent TL diagram.

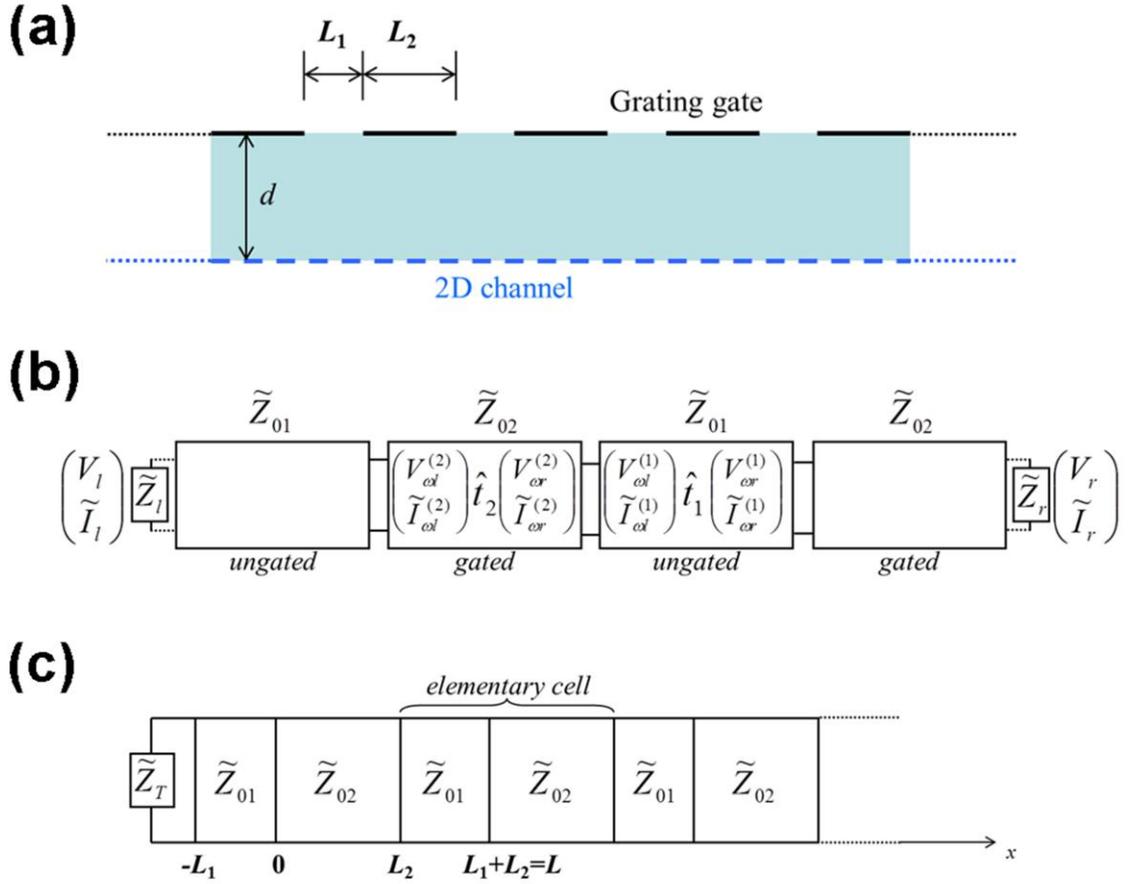

**Figure 2.** (a) 2D electron channel with periodic grating gate and (b) its equivalent segmented TL diagram; (c) segmented TL diagram for semi-infinite plasmonic crystal terminated at one side.

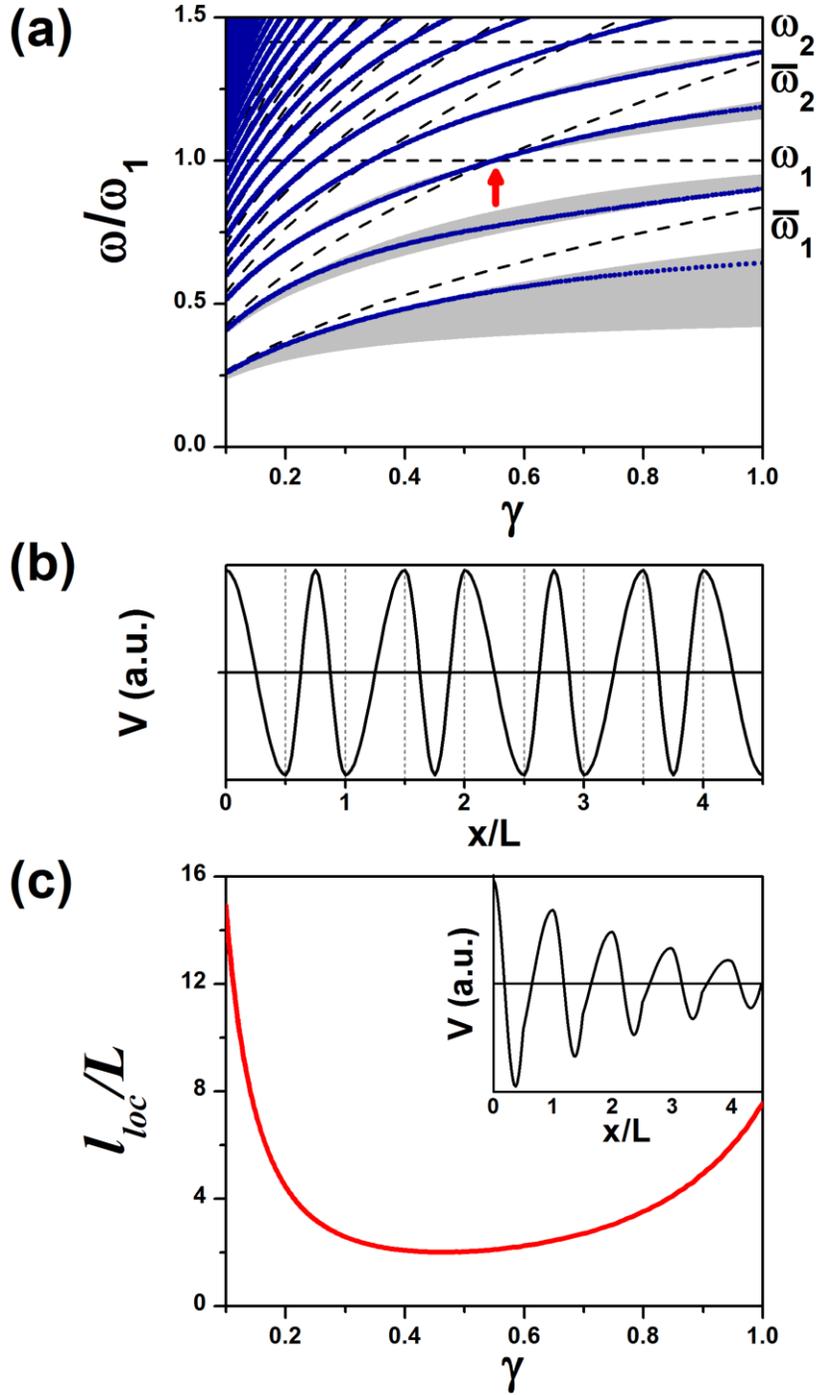

**Figure 3.** (a) The energy band diagram of an infinite 1D plasmonic crystal in the grating gated 2D electron channel as a function of electron density modulation $\gamma$. Band gaps are shown in grey. Dashed lines indicate the energies of the plasma eigenmodes in ungated ($\omega_p$) and gated ($\bar{\omega}_p$) regions. The Tamm states for a semi-infinite crystal are shown by dotted lines. (b) Voltage distribution in the plasma mode at the transparency point indicated by arrow in (a). (c) Localization length of the Tamm state in the second band gap as a function of electron density modulation. Inset: voltage distribution in the Tamm state near the center of the band gap.

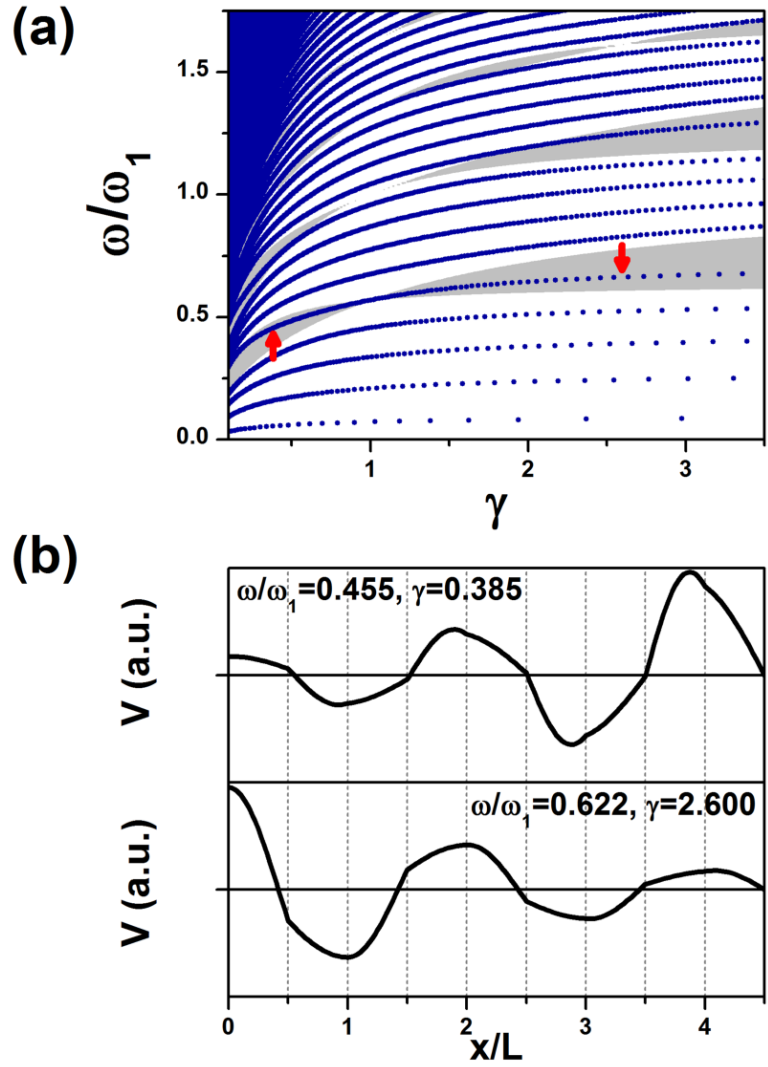

**Figure 4.** (a) The energy band structure of a finite four-period 1D plasmonic crystal terminated by an ohmic contact and an infinite barrier at the opposite boundaries as a function of electron density modulation $\gamma$. Calculated plasma modes are shown by dotted lines. The background shows the energy band spectrum of the corresponding infinite plasmonic crystal. (b) Voltage distributions in the Tamm states calculated at points indicated by arrows in (a).

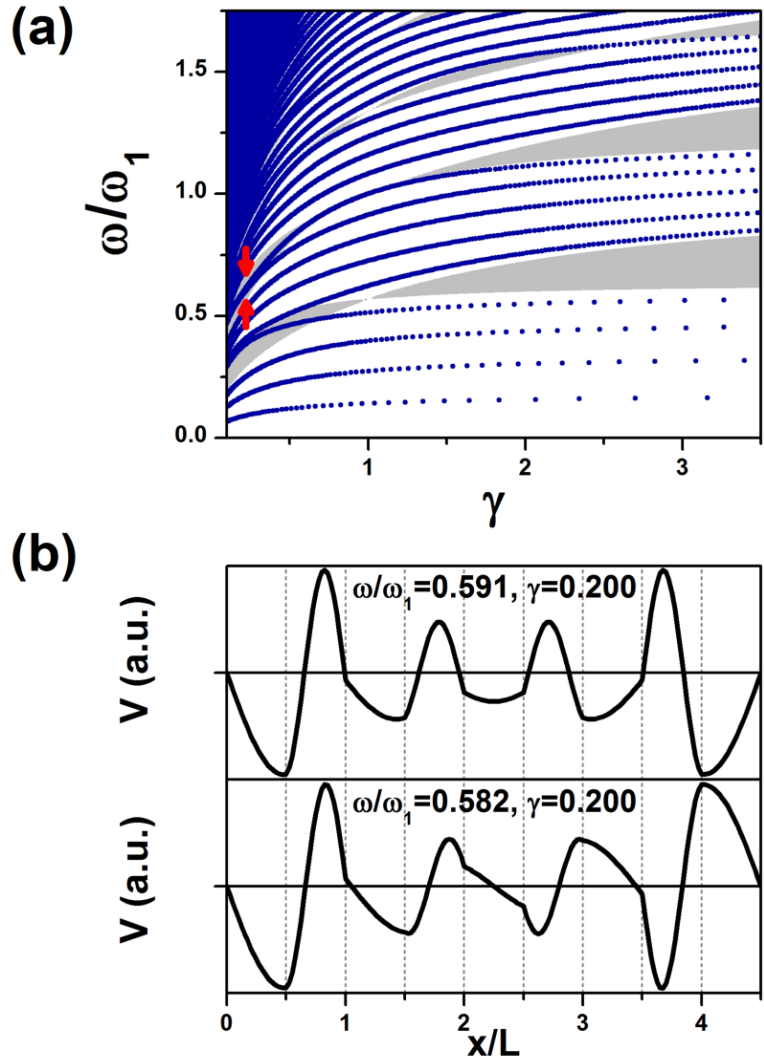

**Figure 5.** (a) The energy band structure of a finite four-period 1D plasmonic crystal terminated by ohmic contacts at both boundaries as a function of electron density modulation $\gamma$. Calculated plasma modes are shown by dotted lines. The background shows the energy band spectrum of the corresponding infinite plasmonic crystal. (b) Voltage distributions in the split symmetric and anti-symmetric Tamm states calculated at points indicated by arrows in (a).

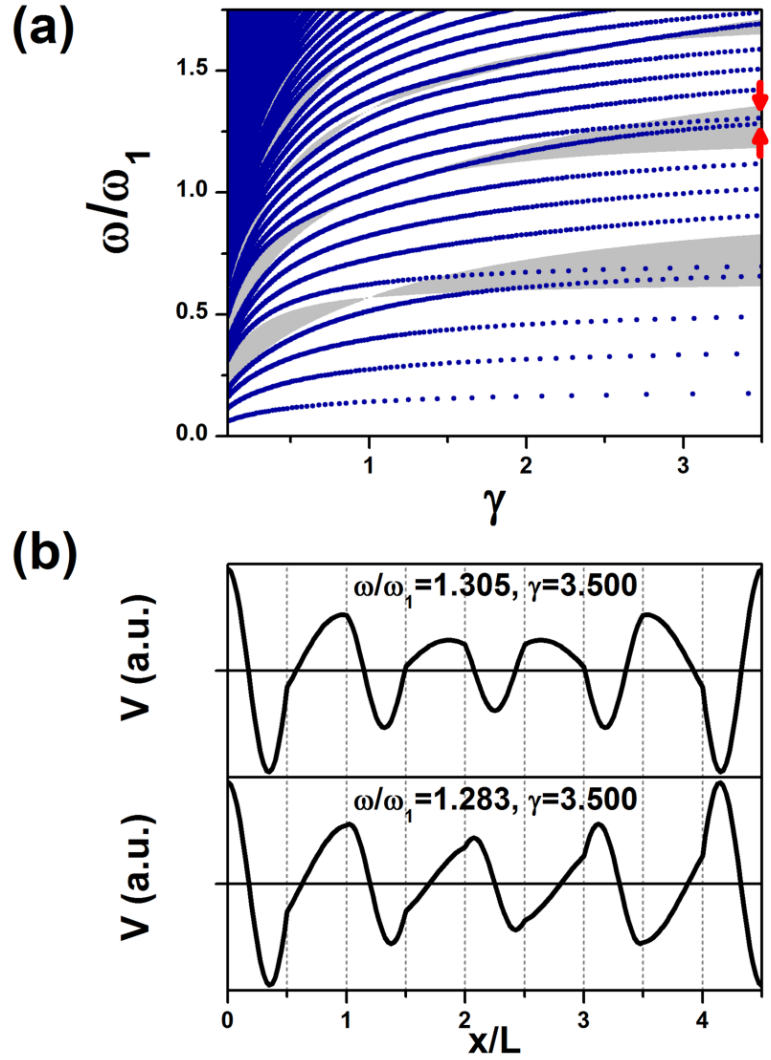

**Figure 6.** (a) The energy band structure of a finite four-period 1D plasmonic crystal terminated by infinite barriers at both boundaries as a function of electron density modulation $\gamma$. Calculated plasma modes are shown by dotted lines. The background shows the energy band spectrum of the corresponding infinite plasmonic crystal. (b) Voltage distributions in the split symmetric and anti-symmetric Tamm states calculated at points indicated by arrows in (a).